\documentclass[twocolumn,a4paper,reprint,amssymb,aps,prd,groupedaddress,superscriptaddress,nofootinbib]{revtex4-1}
\usepackage[dvips]{graphicx}
\usepackage{amssymb}
\usepackage{filecontents}
\usepackage{amsmath}
\usepackage{color}
\usepackage{float}
\usepackage{textcomp}
\usepackage{bm}
\usepackage{lipsum}
\allowdisplaybreaks
\usepackage{mathtools}
\RequirePackage[colorlinks,citecolor=blue,urlcolor=blue,linkcolor=blue]{hyperref}
\usepackage{yfonts}

\usepackage{xcolor}

\begin{document}

\title {Dynamics of the ${N}$-body system in energy-momentum squared gravity: {II}.
  Existence of a Self-Acceleration }

\author{Elham Nazari}
\email{elhamnazari@tafreshu.ac.ir, e.nazari@ipm.ir}
\affiliation{Department of Physics, Faculty of Science, Tafresh University, P. O. Box 39518-79611, Tafresh, Iran}
\affiliation{School of Astronomy, Institute for Research in Fundamental Sciences (IPM), P. O. Box 19395-5531, Tehran, Iran}

\begin{abstract}

We investigate the post-Newtonian (PN) dynamics of energy-momentum squared gravity (EMSG), with particular emphasis on the possibility of self-acceleration in $N$-body systems. A central challenge in matter-type modified gravity theories, including EMSG, is the non-vanishing divergence of the energy-momentum tensor, arising from the nonminimal interaction between the standard and modified matter fields. This feature can, in principle, influence the $N$-body dynamics. In our previous work \cite{2024PhRvD.110f4023N}, its effects on the external-dependent part of the motion were studied in an EMSG class known as quadratic-EMSG. Here, we extend the analysis to the internal-structure-dependent contributions, namely self-acceleration. To this end, we relax the reflection-symmetric assumption adopted in \cite{2024PhRvD.110f4023N} and derive the complete equations of motion for a self-gravitating body in an $N$-body system up to the first PN order. By introducing a suitable expression for the center-of-mass acceleration and employing virial identities\textemdash including one newly emerging within the quadratic-EMSG framework\textemdash it is shown that self-acceleration vanishes. Furthermore, we establish a PN integral conservation law for the total momentum, demonstrating that, as in general relativity (GR), EMSG admits a conserved linear momentum compatible with the absence of self-acceleration. 
Binary pulsar experiments provide stringent bounds on self-acceleration, and our analysis shows that, within the present level of accuracy, EMSG is consistent with these constraints. Therefore, the theory remains viable in the strong-gravity regime probed by binary pulsars.

\end{abstract}
\maketitle

\section{Introduction}

In the previous paper \cite{2024PhRvD.110f4023N}, we studied the $N$-body equations of motion in the post-Newtonian (PN) limit of the energy-momentum squared gravity (EMSG). In this theory, classified as a matter-type modified theory of gravity, the covariant divergence of the energy-momentum tensor $T^{\mu\nu}$ is nonzero, i.e., local conservation of energy and momentum is violated \cite{2014EPJP..129..163K,2016PhRvD..94d4002R,2017PhRvD..96l3517B,2018PhRvD..97b4011A}. It is the result of a nonminimal interaction between the standard and modified matter fields \cite{2024PhRvD.109j4055A}. Beyond this theory, several other modified gravity theories also feature a non-vanishing divergence of the energy-momentum tensor. For instance, see \cite{2001PhRvD..64b4028J,2006PhRvD..73f4015F,2008arXiv0801.1547J} on Einstein-\AE ther theory and \cite{1992CQGra...9.2093D,2003sttg.book.....F} on scalar-tensor theories.
In \cite{1972PhRvD...6.3357R,1975PhRvD..12..376S}, the covariant divergence of the energy-momentum tensor in curved spacetime is questioned, and a class of viable gravitational theories\textemdash modifications of the ordinary Brans-Dicke theory with $\nabla_{\mu}T^{\mu\nu}\neq 0$\textemdash is introduced.

The nonzero divergence of $T^{\mu\nu}$ implies that the trajectories of massive or massless particles may deviate from geodesics. Such deviations can arise from external forces associated with scalar, vector, or tensor fields, from internal unbalanced interactions, or from preferred-frame or preferred-location effects. To investigate this central aspect of the EMSG theory, previous studies have examined the motion of a test particle in the field of a single spherically symmetric body \cite{2022PhRvD.105j4026N,2023PDU....4201305A}, as well as the trajectory of a massive self-gravitating body in the field of an $N$-body system \cite{2024PhRvD.110f4023N}, up to the first PN order. It turns out that, without additional information about the system under study or the parameters of the theory, the EMSG corrections can be absorbed into the definition of the system's mass. As a result, this modified theory cannot be distinguished from general relativity (GR), at least not by Solar System experiments. In terms of the parameters introduced in the parametrized post-Newtonian (PPN) formalism \cite{1971ApJ...163..595T,1971ApJ...163..611W,1971ApJ...169..125W,1974SciAm.231e..24W}, the EMSG theory yields $\gamma=1$ and $\beta=1$. For the current experimental bounds on these PPN parameters, we refer the reader to \cite{2004PhRvL..92l1101S,2008A&A...477..657A,2009IJMPD..18.1129W}. Therefore, with respect to the interpretation of these PPN parameters \footnote{We cautiously adopt these identifications of $\gamma$ and $\beta$ at the level of physical observables.}, the curvature of space and the nonlinearity of the gravitational field in EMSG are, up to the PN order, the same as in GR.

On the other hand, the discussion above does not provide the complete picture, as another potentially measurable EMSG effect can be inferred from the study of the $N$-body equations of motion. To clarify this point, let us highlight an important assumption underlying our previous work: each body in the $N$-body system was assumed to be reflection-symmetric about its center of mass \cite{2024PhRvD.110f4023N}. In the Solar System, where bodies are highly symmetric and orbits are nearly circular, this assumption is well justified, since asymmetric effects are utterly negligible. However, such symmetries \textit{do} automatically eliminate a significant contribution to the $N$-body equations of motion\textemdash namely, the self-acceleration of the body. In general, the motion of a body's center of mass depends not only on external fields but also on its internal structure and composition, giving rise to self-acceleration \cite{will2018theory}. While self-acceleration is absent in GR, even in asymmetric cases, it can emerge in non-GR theories, particularly when $T^{\mu\nu}$ is not divergence-free. The condition  $\nabla_{\mu}T^{\mu\nu}\neq 0$ thus sets the stage for the possibility of self-acceleration in modified gravity theories. This component of the $N$-body acceleration is especially relevant in the dynamics of asymmetric systems, with binary pulsars in eccentric orbits serving as a prime example.  
As high-precision laboratories, binary pulsars impose stringent constraints on gravitational theories in the strong-field gravity regime and provide a critical benchmark for their viability (see \cite{2021PhRvX..11d1050K} and references therein). Self-acceleration can also be tested with high accuracy in these systems \cite{1976ApJ...205..861W,1992ApJ...393L..59W,2020ApJ...898...69M}. Consequently, if a gravity theory predicts self-acceleration in a binary system, its free parameters must be tightly constrained to remain consistent with binary pulsar observations.
The EMSG theory has already been confronted with several gravitational experiments in both weak and strong-field regimes   \cite{2018PhRvD..98b4031N,2018PhRvD..97l4017A,2022PhRvD.105d4014N,2022PhRvD.105j4026N,2023PDU....4201305A,2023MNRAS.523.5452A,2023PDU....4201360M,2023JCAP...08..067M,2024PhRvD.110f4023N,2024PDU....4501505A}. Motivated by the unparalleled precision of binary pulsar observations, in this paper we study the complete $N$-body equations of motion in the EMSG framework, with a particular focus on the possibility of self-acceleration.

The central aim of this work is to relax the aforementioned assumption of reflection symmetry, thereby capturing the full range of possible EMSG effects in $N$-body dynamics and enabling the study of eccentric binary systems within this modified gravity framework.
Among the several classes of EMSG theories \cite{2014EPJP..129..163K,2016PhRvD..94d4002R,2017PhRvD..96l3517B,2018PhRvD..97b4011A,2018PhRvD..98f3522A,2019EPJC...79..846A,2020PhRvD.102f4016N,2023MNRAS.523.5452A}, we concentrate on the quadratic-EMSG category, which incorporates second-order matter contributions into the field equations. If self-acceleration arises, it will manifest in the orbital dynamics, a subject we investigate up to the first PN order. Accordingly, the main purpose of the present work is to examine the PN consistency of the theory at the level of N-body dynamics, independent of the separate question of its full cosmological viability. Appendix \ref{app1} summarizes the PN limit of this theory and provides the foundation for the subsequent analysis. In Sec. \ref{Sec. II}, we present the conventional quadratic-EMSG field equations, and derive the general form of the $N$-body equations of motion, incorporating all relevant EMSG effects. 
Sec. \ref{Sec. III} is devoted to the extraction of self-acceleration, for which we introduce virial identities in this modified theory. The derivation in this section is conceptually straightforward but technically lengthy, so the detailed steps are deferred to Appendix \ref{app2}.
Since self-acceleration is intrinsically linked to the breakdown of momentum conservation \cite{will2018theory}, we dedicate Sec. \ref{Sec. IV} to this issue in the quadratic-EMSG setting. Finally, Sec. \ref{Sec. V} provides an overview of our results and draws the conclusions.

Throughout this work, we adopt the conventions of our previous study \cite{2024PhRvD.110f4023N}, to which readers are referred for additional background, concepts, and definitions.

\section{Motion of isolated bodies in Energy-momentum squared gravity} \label{Sec. II}
Initially, our discussion focuses on the introduction of the quadratic-EMSG theory. 
The conventional form of the matter-type modified gravity theories has recently been analyzed in \cite{2024PhRvD.109j4055A}. In these theories, the second metric derivative of the matter Lagrangian density, $\mathcal{L}_{\text{m}}$, is assumed to vanish. Upon careful examination, it becomes apparent that using the standard formulation confines the analysis to the level of the field equations, effectively precluding a full action/Lagrangian formulation for this class of modified theories.  
Accordingly, we first introduce the conventional quadratic-EMSG field equations and then proceed to the detailed derivation of the $N$-body dynamics.

\subsection{Energy-momentum squared gravity}

The quadratic-EMSG field equations are introduced in \cite{2016PhRvD..94d4002R}. According to the notation used in \cite{2024PhRvD.110f4023N}, they are expressed as follows:
\begin{align}\label{G}
G_{\mu\nu}=
k\Big( T_{\mu\nu}+f_0'\big( g_{\mu\nu}{\boldsymbol{T}}^2-4T^\sigma_\mu T_{\nu\sigma}-4{\boldsymbol{\Psi}}_{\mu\nu}\big)\Big),
\end{align}
where $k=8\pi G/c^4$, $f_0'$ is the free parameter of the theory, $g_{\mu\nu}$ is the spacetime metric with the determinant $g$, $G_{\mu\nu}=R_{\mu\nu}-\frac{1}{2}g_{\mu\nu}R$ is the Einstein tensor, where $R_{\mu\nu}$ and $R$ are the Ricci tensor and the Ricci scalar, respectively, and   
\begin{align}
{\boldsymbol{\Psi}}_{\mu\nu}=-\mathcal{L}_{\text{m}}\big(T_{\mu\nu}-&\frac{1}{2}Tg_{\mu\nu}\big)\\\nonumber
&-\frac{1}{2}TT_{\mu\nu}-2T^{\alpha\beta}\frac{\partial^2 \mathcal{L}_{\text{m}}}{\partial g^{\alpha\beta}\partial g^{\mu\nu}}.
\end{align}
Here, $\mathcal{L}_{\text{m}}$ is the matter Lagrangian density, $T$ is the trace of the energy-momentum tensor $T^{\mu\nu}$, and $\boldsymbol{T}^2=T^{\mu\nu}T_{\mu\nu}$.
It should be mentioned that the matter Lagrangian density is independent of metric derivatives.

Using the Bianchi identities, we get 
\begin{align}\label{nabla_T_eff}
\nabla_{\mu}T_{\text{eff}}^{\mu\nu}=0.
\end{align} 
$T^{\text{eff}}_{\mu\nu}$ is the effective conserved energy-momentum tensor defined as
\begin{align}\label{T_eff}
T^{\text{eff}}_{\mu\nu}=T_{\mu\nu}+T^{\text{\tiny EMSG}}_{\mu\nu},
\end{align}
in which  $T_{\mu\nu}$ is the standard portion and 
\begin{align}
T^{\text{\tiny EMSG}}_{\mu\nu}=f'_0\Big( g_{\mu\nu}{\boldsymbol{T}}^2-4T^\sigma_\mu T_{\nu\sigma}-4{\boldsymbol{\Psi}}_{\mu\nu}\Big), 
\end{align}
is the EMSG part. Rewriting Eq.~\eqref{nabla_T_eff} yields
\begin{align}
\nabla_{\mu}T^{\mu\nu}=-\nabla_{\mu}T_{\text{\tiny EMSG}}^{\mu\nu},
\end{align} 
which indicates that the standard matter does not conserve energy and momentum independently; any momentum lost or gained by the standard part is exchanged with the EMSG field. In fact, in this theory, the usual stressed matter is represented by the energy-momentum tensor, which, unlike in GR, is not divergence-free. Hereafter, we assume that the standard matter is a perfect fluid described by $\mathcal{L}_{\text{m}}=p$ \cite{1970PhRvD...2.2762S,1993CQGra..10.1579B} and consistently omit the last term in the definition of ${\boldsymbol{\Psi}}_{\mu\nu}$ \cite{2024PhRvD.109j4055A}.   

Closely patterned after our previous works \cite{2024PhRvD.110f4023N,2022PhRvD.105d4014N,2022PhRvD.105j4026N,2023MNRAS.523.5452A}, we consider that, in this class of modified theories, baryon number is unambiguously conserved, assured by the continuity equation
\begin{align}
\nabla_{\mu}\big(\rho\, u^{\mu}\big)=0.
\end{align}
Here, $\rho$ is the rest-mass density, $u^{\mu}=\gamma(c,\boldsymbol{v})$ is the four-velocity, where $\boldsymbol{v}$ is the three-velocity field, $c$ is the speed of light, and $\gamma=u^0/c$.   
This equation illustrates that the rescaled mass density, defined as $\rho^*=\sqrt{-g}\gamma \rho$, is conserved:
\begin{align}\label{rhos}
\partial_t\rho^*+\partial_j\big(\rho^* v^{j}\big)=0.
\end{align}
This relation, together with the redefinition of density, provides a powerful and frequently used tool for simplifying the calculations presented below.

\subsection{Center-of-mass variables}

To introduce the $N$-body dynamics, we begin with the definitions of the center-of-mass variables that characterize the motion of a body in an $N$-body system.
As a starting point, we consider the definition of total inertial mass in general spacetimes. 
Unlike Newtonian gravity, no coordinate-invariant definition of total inertial mass exists in curved spacetime. Nevertheless, for systems in asymptotically flat spacetimes, several notions of mass have been developed, each useful in different contexts\textemdash for example, the ADM (Arnowitt-Deser-Misner) mass \cite{2008GReGr..40.1997A}, the Komar mass \cite{1984ucp..book.....W,2004rtmb.book.....P}, and the Bondi mass \cite{1962PhRv..128.2851S}. In addition, the inertial mass can be expressed as the spatial integral of the energy density \cite{will2018theory}.   
Building on this point, we introduce the inertial mass in the EMSG framework through the total mass-energy.

In \cite{2024PhRvD.110f4023N}, by examining the temporal component of Eq.~\eqref{nabla_T_eff}, the total mass-energy of a body, hereafter denoted by $A$, is shown to be
\begin{align}
M_A\equiv m_A+\frac{E_A}{c^2},
\end{align} 
where $m_A=\int_A\rho^*\,d^3x$, the integral being taken over the volume occupied by body $A$, and $E_A$ denotes its total energy, given by 
\begin{align}\label{total_energy}
E_A=\mathcal{T}_A+\Omega_A+E^{\text{int}}_A+f_0'c^4\mathfrak{M}_A+O(c^{-2}).
\end{align}
Here, $\mathcal{T}_A$ is the kinetic energy, $\Omega_A$ is the gravitational potential energy, $E^{\text{int}}_A$ is the internal energy, and $\mathfrak{M}_A$ represents the EMSG contribution to the body's energy. The definitions of these scalar quantities are provided in Appendix \ref{app2}; see \eqref{eq1}\textendash\eqref{eq4}.  
Given these definitions, the integral form of the total mass-energy of body $A$ is written as  
\begin{align}\label{M_A}
\nonumber
M_A\equiv&\int_A\rho^*\Big[1+\frac{1}{c^2}\big(\Pi+\frac{1}{2}\bar{v}^2-\frac{1}{2}U_A+f_0' c^4\rho^*\big)\Big]d^3x\\
&+O(c^{-4}),
\end{align}
where $U_A$ is the internal part of the standard potential, $U=U_A+U_{-A}$, with
\begin{subequations}
\begin{align}
\label{U_A}
& U_A=G\int_{A}\frac{{\rho^*}'}{\rvert{\boldsymbol{x}-\boldsymbol{x}'}\rvert}d^3x',\\
\label{U__A}
& U_{-A}=\sum_{B\neq A}G\int_{B}\frac{{\rho^*}'}{\rvert{\boldsymbol{x}-\boldsymbol{x}'}\rvert}d^3x'.
\end{align}
\end{subequations}
Here, $\Pi$ is the internal energy of a fluid element divided by its mass and $\bar{\boldsymbol{v}}=\boldsymbol{v}-\boldsymbol{v}_{A(0)}(t)$ is its velocity relative to the velocity of body $A$, $\boldsymbol{v}_{A(0)}$. Another useful relative vector is $\bar{\boldsymbol{x}}=\boldsymbol{x}-\boldsymbol{r}_{A(0)}(t)$, which measures the position of the element with respect to the center of mass of body $A$, located at $\boldsymbol{r}_{A(0)}$. The precise definitions of $\boldsymbol{r}_{A(0)}$ and $\boldsymbol{v}_{A(0)}$ are given below. 

According to Eq.~\eqref{M_A}, the center-of-mass position of body $A$ is formulated as follows:
\begin{align}\label{R_A}
\nonumber
{R}_A^j\equiv&\frac{1}{M_A}\int_A\rho^*{x}^j\Big[1+\frac{1}{c^2}\big(\Pi+\frac{1}{2}\bar{v}^2-\frac{1}{2}U_A\\
&+f_0'c^4\rho^*\big)\Big]d^3x+O(c^{-4}),
\end{align}
whose lowest-order term is represented by $\boldsymbol{r}_{A(0)}$. 
Given that $dM_A/dt=0$ \cite{2024PhRvD.110f4023N}, and introducing $\boldsymbol{v}_A\equiv d\boldsymbol{R}_A/dt$, we obtain
\begin{align}\label{v_A}
&
\nonumber v_A^j\equiv\frac{1}{M_A}\bigg\lbrace\int_A\Big[\rho^*v^j\Big(1+\frac{1}{c^2}\big(\Pi+\frac{1}{2}\bar{v}^2-\frac{1}{2}U_A+f_0'c^4\rho^*\big)\Big)\\
&+\frac{1}{c^2}\Big(p\bar{v}^j-\frac{1}{2}\rho^*\bar{W}_A^j+f_0'c^4{\rho^*}^2\bar{v}^j\Big)\Big]d^3x\bigg\rbrace+O(c^{-4}),
\end{align}
where
\begin{align}
\bar{W}_A^j=G\int_A{\rho^*}'\bar{v}'_k\frac{\big(x-x'\big)^k\big(x-x'\big)^j}{\rvert{\boldsymbol{x}-\boldsymbol{x}'}\rvert^3}d^3x'.
\end{align}
To simplify Eq.~\eqref{v_A}, we employ Eq.~\eqref{rhos}, the Newtonian order of Eq.~\eqref{Euler-PN-EMSG}, Eq.~\eqref{first-law}, and the integral identities presented in Appendix C of \cite{2024PhRvD.110f4023N}. We also make use of the fact that both the pressure and density vanish at the surface of the bodies. The quantity $\boldsymbol{v}_{A(0)}$ denotes the leading-order contribution to the velocity of the body's center of mass, as given in Eq.~\eqref{v_A}.  

Finally, considering the definition $\boldsymbol{a}_A\equiv d\boldsymbol{v}_A/dt$, we arrive at
\begin{align}\label{a_j-EMSG}
\nonumber
& a_A^j=\frac{1}{M_A}\bigg\lbrace\int_A\rho^*\Big[1+\frac{1}{c^2}\big(\Pi+\frac{1}{2}\bar{v}^2-\frac{1}{2}U_A+f_0'c^4\rho^*\big)\Big]\frac{dv^j}{dt}d^3x\\\nonumber
&+\frac{1}{c^2}\Big[\mathcal{P}^j_A+\frac{1}{2}\big(\mathcal{T}^j_A-{\mathcal{T}^*}^j_A\big)+\int_A\big(\partial_tp\,\bar{v}^j-\frac{p}{\rho^*}\partial_jp\big)d^3x\\\nonumber
&+v_A^k\int_A\bar{v}^j\partial_kp\,d^3x-\frac{1}{2}\frac{d}{dt}\int_A\rho^*\bar{W}^j_A\,d^3x\Big]+f_0'c^2\Big[{\Omega^{**}_A}^j\\\nonumber
&+v_A^j\frac{d}{dt}\mathfrak{M}_A+\int_A\rho^*\,\partial_jp\,d^3x-2\int_A{\rho^*}^2\bar{v}^j\partial_k\bar{v}^k\,d^3x\\
&-2\int_A\rho^*v^j\bar{v}^k\partial_k\rho^*d^3x\Big]\bigg\rbrace+O(c^{-4}).
\end{align}
This expression represents the complete form of the equations of motion of isolated gravitating bodies in the quadratic-EMSG theory. 
The vector quantities $\boldsymbol{\mathcal{T}}_A$, $\boldsymbol{\mathcal{T}}^*_A$, $\boldsymbol{\Omega}^{**}_A$, and $\boldsymbol{\mathcal{P}}_A$, which pertain to the internal structure of the $A$th body, are defined in Eqs.~\eqref{T_A}, \eqref{T*_A}, \eqref{Omega**_A}, and \eqref{P_A}, respectively. In the absence of the EMSG corrections, this equation reduces to the standard expression for the acceleration of the body's center of mass given in \cite{will2018theory} (see Eq.~(6.39) therein). Under the additional assumption of reflection symmetry about the origin of the body, Eq.~\eqref{a_j-EMSG} further simplifies to its first term, i.e., $\int_A\rho^*\frac{dv^j}{dt}d^3x$, at the first PN order of the $N$-body dynamics, as discussed in detail in \cite{2024PhRvD.110f4023N}. In the present work, however, we relax this symmetry assumption to investigate the effect of self-acceleration on the body's motion, which leads to a more complex relation. The analysis of Eq.~\eqref{a_j-EMSG} forms the focus of the following section.

\section{Self acceleration in Energy-momentum squared gravity} \label{Sec. III}
First, by means of a crude estimation, we show that the EMSG contribution can modify the $N$-body dynamics, in particular through the component determined purely by the body's internal structure. Subsequently, we examine all the terms in Eq.~\eqref{a_j-EMSG} to identify the role of the EMSG corrections in self-acceleration.

\subsection{Crude estimation}

In the previous section, we obtained that $\boldsymbol{a}_A\propto\int_A\rho^*\big(d\boldsymbol{v}/dt\big)d^3x$ at the lowest order. 
From the perspective of fluid dynamics (see Eq.~\eqref{Euler-PN-EMSG}), several EMSG corrections contribute to $d\boldsymbol{v}/dt$ at both Newtonian and PN orders. Following a similar approach to \cite{2024PhRvD.110f4023N}, we adopt the Newtonian Euler equation, supplemented by a relativistic contribution from the EMSG terms, i.e., $4f_0'c^2\rho^* \partial_jU_{\text{\tiny EMSG}}$. Thus, we have
\begin{align}
\rho^*\frac{dv^j}{dt}\approx\rho^*\partial_jU-\partial_jp-2f_0'c^4\rho^*\Big(\partial_j\rho^*-\frac{2}{c^2}\partial_jU_{\text{\tiny EMSG}}\Big).
\end{align}
Substituting this relation into the definition of $\boldsymbol{a}_A$ yields 
\begin{align}\label{a-estimate}
&\int_A\rho^*\big(dv^j/dt\big)d^3x=\int_A\rho^*\Big[\partial_jU_A+\partial_jU_{-A}\\\nonumber
&+4f_0'c^2\big(\partial_jU_{\text{\tiny EMSG}, A}+\partial_jU_{\text{\tiny EMSG}, -A}\big)\Big]d^3x,
\end{align} 
in which the gravitational potentials are separated into two parts; see the third assumption mentioned in Appendix \ref{app1}. Specifically, $U_A$ and $U_{-A}$ denote the internal and external pieces of $U$, as defined in Eqs.~\eqref{U_A} and \eqref{U__A}, respectively. Similarly, 
\begin{subequations}
\begin{align}
\label{U_A_EMAG}
& U_{\text{\tiny EMSG},A}=G\int_{A}\frac{{{\rho^*}'}^2}{\rvert\boldsymbol{x}-\boldsymbol{x}'\rvert}\,d^3x',\\
\label{U__A_EMSG}
& U_{\text{\tiny EMSG},-A}=\sum_{B\neq A}G\int_{B}\frac{{{\rho^*}'}^2}{\rvert\boldsymbol{x}-\boldsymbol{x}'\rvert}\,d^3x',
\end{align}
\end{subequations}
represent the internal and external pieces of the EMSG gravitational potential $U_{\text{\tiny EMSG}}$, respectively. By inserting Eqs.~\eqref{U_A}\textendash\eqref{U__A} and \eqref{U_A_EMAG}\textendash\eqref{U__A_EMSG} in Eq.~\eqref{a-estimate}, using the definition of the relative vector $\bar{\boldsymbol{x}}$, and after interchanging the integration variables $\bar{\boldsymbol{x}}\leftrightarrow \bar{\boldsymbol{x}}'$, one finds that 
\begin{align}
a_A^j\propto \int_A\rho^*\Big[\partial_jU_{-A}+4f_0'c^2\big(\partial_jU_{\text{\tiny EMSG}, A}+\partial_jU_{\text{\tiny EMSG}, -A}\big)\Big]d^3x.
\end{align}
As seen, the internal part of $U$ does not contribute to the body's motion, whereas $U_{\text{\tiny EMSG},A}$, which depends solely on the body's internal structure, can change it. This EMSG term may therefore induce a self-acceleration, $a_{\text{self}}^j$, in this theory. 
It should be noted that imposing the symmetry condition on the bodies causes this term to vanish automatically. This estimation also suggests that $U_{\text{\tiny EMSG},-A}$ may play a role in the external part of the equations of motion, indicated by $a^j_{\text{N-body}}$, which was the focus of our previous work \cite{2024PhRvD.110f4023N}. A closer inspection of Eq.~\eqref{a_j-EMSG} further reveals that additional EMSG terms, such as the vector quantity $\boldsymbol{\Omega}^{**}_A$, may also induce self-acceleration in the body's motion.

\subsection{Equations of motion }

To complete this analysis, we investigate each term in Eq.~\eqref{a_j-EMSG}.
To do so, we employ the PN EMSG Euler equations \eqref{Euler-PN-EMSG} for the first part of Eq.~\eqref{a_j-EMSG}, namely $\int_A\rho^*\frac{dv^j}{dt}d^3x$, while the remaining PN-order parts are simplified using the Newtonian EMSG Euler equations.  
For convenience in later calculations, we introduce the auxiliary potentials $\Phi_1$\textendash$\Phi_6$ (cf. Eqs.~\eqref{Phi_1}\textendash\eqref{Phi_6}), which allow us to rewrite $\psi+V= \frac{3}{2}\Phi_1-\Phi_2+\Phi_3+3\Phi_4$ and $\partial_{tt}X=\Phi_1+2\Phi_4-\Phi_5-\Phi_6+2f_0'c^4U_{\text{\tiny{EMSG}}}$. Since $\partial_{tt}X$ is a PN correction, it is derived using the Newtonian EMSG Euler equations. In this notation, the potential $\Psi$ appearing in Eq.~\eqref{Euler-PN-EMSG} can be expressed as  $\Psi=2\Phi_1-\Phi_2+\Phi_3+4\Phi_4-\frac{1}{2}\Phi_5-\frac{1}{2}\Phi_6+f_0'c^4U_{\text{\tiny{EMSG}}}$ \cite{2024PhRvD.110f4023N}.

To address the complex expression for $\boldsymbol{a}_A$, we decompose the first long integral into smaller, more manageable components, referred to as force integrals. This is given by 
\begin{align}\label{frist-term}
\int_A\rho^*\frac{dv^j}{dt}d^3x=F_0^j+\sum_{n=1}^{30}F^j_n+O(c^{-4}),
\end{align} 
where $F_0^j\textendash F_{30}^j$ are defined in Appendix \ref{app2}. 
For the next part of Eq.~\eqref{a_j-EMSG}, we arrive at
\begin{align}\label{second-term}
\nonumber
&\frac{1}{c^2}\int_A\rho^*\Big(\Pi+\frac{1}{2}\bar{v}^2-\frac{1}{2}U_A+f_0'c^4\rho^*\Big)\frac{dv^j}{dt}d^3x\\
&=-F_3^j+F_{21}^j-\frac{1}{2}F_{25}^j+\frac{1}{6}F_{28}^j+\sum_{n=31}^{38}F^j_{n}+O(c^{-4}),
\end{align}
in which $F_{31}^j\textendash F_{38}^j$ are also introduced in Appendix. \ref{app2}.

By applying the first, third, and fourth assumptions outlined in Appendix \ref{app1}, and using the relative vectors $\bar{\boldsymbol{x}}$ and $\bar{\boldsymbol{v}}$, together with the identity $\int_A\rho^*\bar{v}^k\partial_k\rho^*\,d^3x=\frac{1}{2}\frac{d}{dt}\mathfrak{M}_A$, the force integrals $F_{0}^j\textendash F_{38}^j$ are simplified, cf. Eqs.~\eqref{F_0}\textendash\eqref{F_38}.
Among these force integrals, $F_{19}^j\textendash F_{30}^j$, $F_{36}^j\textendash F_{38}^j$, as well as a part of $F_{12}^j$ arise from the EMSG field.    
Substituting these results into Eqs.~\eqref{frist-term} and \eqref{second-term}, subsequently into Eq.~\eqref{a_j-EMSG}, we obtain, after some manipulations,
\begin{widetext}
\begin{align}\label{a_A_b}
\nonumber
&M_A\,a_A^j=m_A\partial_jU_{-A}+\frac{1}{c^2}\bigg\lbrace v^k_A\Big[2L_A^{(jk)}+3K_A^{jk}-4H_A^{(jk)}-\delta^{jk}\frac{d}{dt}P_A+2f_0'c^4\big(2Q_A^{(jk)}-\delta^{jk}Q_A\big)\Big]-\frac{1}{2}\Big[t^j_A+\mathcal{P}^j_A-{\mathcal{T}^{*}_A}^j\\\nonumber
&+\mathcal{T}^j_A-3{\mathcal{T}^{**}_A}^j+{\Omega^*_A}^j-3f_0'c^4{\Omega^{**}_A}^j\Big]+\partial_jU_{-A}\Big[3P_A+2\mathcal{T}_A+\Omega_A+E_A+3f_0'c^4\mathfrak{M}_A\Big]-4\partial_kU_{-A}\Big[\Omega^{jk}_A+2\mathcal{T}^{jk}_A+\delta^{jk}P_A\\\nonumber
&+f_0'c^4\delta^{jk}\mathfrak{M}_A\Big]-\frac{1}{2}\frac{d}{dt}\int_A\rho^*\bar{W}^j_A\,d^3x+2f_0'c^4\Big(\int_Ap\,\partial_j\rho^*d^3x-\frac{d}{dt}\mathfrak{P}_A^j-\int_A{\rho^*}^2\bar{v}^j\partial_k\bar{v}^kd^3x\Big)\bigg\rbrace+\frac{m_A}{c^2}\bigg\lbrace\partial_jU_{-A}\Big(v_A^2-4U_{-A}\Big)\\\nonumber
&-v_A^j\Big(3\partial_tU_{-A}+4v_A^k\partial_kU_{-A}\Big)-4v_A^k\Big(\partial_jU_{k,-A}-\partial_kU_{j,-A}\Big)+4\partial_tU_{j,-A}+2\partial_j\Phi_{1,-A}-\partial_j\Phi_{2,-A}+\partial_j\Phi_{3,-A}+4\partial_j\Phi_{4,-A}\\
&-\frac{1}{2}\partial_j\Phi_{5,-A}-\frac{1}{2}\partial_j\Phi_{6,-A}+5f_0'c^4\partial_jU_{\text{\tiny EMSG},-A}\bigg\rbrace+O(c^{-4}).
\end{align}
\end{widetext} 
Parentheses around indices denote symmetrization; for example, $T^{(jk)}=\frac{1}{2}\big(T^{jk}+T^{kj}\big)$. The new scalar, vector, and tensor quantities in the above relation are defined in Appendix \ref{app1}. Here, we apply the identity $\Lambda_A^j=-\frac{1}{2}{\Omega^{**}_A}^j$ to simplify this relation.    
Since $M_A=m_A+\frac{E_A}{c^2}$, the Newtonian term $m_A\partial_jU_{-A}$ can be combined with the PN part $\frac{E_A}{c^2}\partial_jU_{-A}$ in the second line, yielding $M_A\partial_jU_{-A}$. For the remaining PN terms, $m_A$ can simply be replaced by $M_A$.

In comparison with Eq.~(33) of \cite{2024PhRvD.110f4023N}, it becomes evident that several additional vector quantities\textemdash most prominently the EMSG-induced terms arising purely from the body's internal structure\textemdash enter the acceleration equation. These contributions, which have no analogue in the standard PN framework of GR, represent genuinely novel effects that may influence the $N$-body dynamics and, in particular, can give rise to self-acceleration of bodies. 
To complete the analysis, we impose the second assumption from Appendix \ref{app1}, namely the internal dynamical equilibrium of body $A$, to further simplify the above result. For this purpose, we introduce the virial identities in the following subsection.

\subsection{Virial identities}
 
In \cite{2024PhRvD.110f4023N}, we derived the virial identities constructed from the total time derivatives of the quadrupole-moment tensor $I^{jk}$, Eq.~\eqref{I^jk}, in EMSG.  For convenience, they are listed below.  
\begin{subequations}
\begin{align}
\label{vir_1}
&\frac{1}{2}\frac{d}{dt}{I}^{jk}_A=\frac{1}{2}S_A^{jk}+\int_A\rho^*\bar{x}^k\bar{v}^j\,d^3x,\\
&\frac{1}{2}\frac{d^2}{dt^2}{I}^{jk}_A=2\mathcal{T}^{jk}_A+\Omega^{jk}_A+\delta^{jk}P_A+f_0'c^4\delta^{jk}\mathfrak{M}_A+O(c^{-2}),\\\nonumber
\label{vir_3}
&\frac{1}{2}\frac{d^3}{dt^3}{I}^{jk}=4H^{(jk)}_A-2L_A^{(jk)}+\delta^{jk}\frac{d}{dt}{P}_A-3K_A^{(jk)}\\
&~~~~~~~~~~~~~~-f_0'c^4\big(4Q^{(jk)}_A-\delta^{jk}\frac{d}{dt}\mathfrak{M}_A\big)+O(c^{-2}).
\end{align}
\end{subequations}
Here, the spin tensor $S_A^{jk}$ in the first identity vanishes for a non-spinning body; For details, the reader is referred to Subsec. III.C of \cite{2024PhRvD.110f4023N}. Another relevant vectorial identity is obtained from the total time derivative of the internal-structure-dependent quantity $\int_A\rho^*\bar{W}^j_A\,d^3x$. After some algebra, we find   
\begin{align}\label{vir_4}
\nonumber
&\frac{d}{dt}\int_A\rho^*\bar{W}^j_A\,d^3x=-{\Omega^*_A}^j-\mathcal{P}^j_A-t^j_A-\mathcal{T}^j_A+{\mathcal{T}^*_A}^j\\
&~~~~~~~~~~~~~~~~~~~~~~~+3{\mathcal{T}^{**}_A}^j-f_0'c^4{\Omega_A^{**}}^j+O(c^{-2}),
\end{align}
in which the Newtonian EMSG Euler equation is employed. 
It is worth noting that these results are retained only to the PN order required to simplify the equations of motion. 

After making use of the virial identities \eqref{vir_1}\textendash\eqref{vir_3} and \eqref{vir_4}, we derive the equilibrium conditions:
\begin{subequations}
\begin{align}
\label{condi_1}
0=&\int_A\rho^*\bar{x}^k\bar{v}^j\,d^3x,\\
0=
\label{cond_2} &~2\mathcal{T}^{jk}_A+\Omega^{jk}_A+\delta^{jk}P_A+f_0'c^4\delta^{jk}\mathfrak{M}_A+O(c^{-2}),\\\nonumber
0= &~4H^{(jk)}_A-2L_A^{(jk)}+\delta^{jk}\frac{d}{dt}{P}_A-3K_A^{(jk)}-f_0'c^4\big(4Q^{(jk)}_A\\
&-\delta^{jk}\frac{d}{dt}\mathfrak{M}_A\big)+O(c^{-2}),\\\nonumber
0= &-{\Omega^*_A}^j-\mathcal{P}^j_A-t^j_A-\mathcal{T}^j_A+{\mathcal{T}^*_A}^j+3{\mathcal{T}^{**}_A}^j-f_0'c^4{\Omega_A^{**}}^j\\\label{condi_4}
&+O(c^{-2}),
\end{align}
\end{subequations}
where the total time derivatives of all internal quantities are set to zero, reflecting the internal dynamical equilibrium of the body. Furthermore, to complete this result, we consider the trace of Eq.~\eqref{cond_2} as an additional equilibrium condition, namely 
\begin{align}
\label{condi_5}
0=2\mathcal{T}_A+\Omega_A+3P_A+3f_0'c^4\mathfrak{M}_A+O(c^{-2}).
\end{align}

\subsection{Final result}
We are now in a position to finalize the derivation. Substituting the conditions \eqref{condi_1}\textendash\eqref{condi_4} and \eqref{condi_5} into Eq.~\eqref{a_A_b}, we deduce that 
\begin{align}\label{a_A_f}
a^j_A=\big(a^j_A\big)_{\text{Newt}}+\big(a^j_A\big)_{\text{self}}+\big(a^j_A\big)_{\text{N-body}},
\end{align}
where
\begin{subequations}
\begin{align}
&\big(a^j_A\big)_{\text{Newt}}=\partial_jU_{-A},\\\nonumber
\label{a_self}
&\big(a^j_A\big)_{\text{self}}=\frac{2f_0'c^2}{M_A}\bigg\lbrace{\Omega_A^{**}}^j+\int_A p\,\partial_j\rho^*\,d^3x\\
&~~~~~~~~~~~~~~~~~~~~~-\int_A{\rho^*}^2\bar{v}^j\partial_k\bar{v}^k\,d^3x\bigg\rbrace+O(c^{-4}),\\\nonumber
&\big(a^j_A\big)_{\text{N-body}}=\frac{1}{c^2}\bigg\lbrace\partial_jU_{-A}\Big(v_A^2-4U_{-A}\Big)-v_A^j\Big(3\partial_tU_{-A}\\\nonumber
&+4v_A^k\partial_kU_{-A}\Big)-4v_A^k\Big(\partial_jU_{k,-A}-\partial_kU_{j,-A}\Big)+4\partial_tU_{j,-A}\\\nonumber
&+2\partial_j\Phi_{1,-A}-\partial_j\Phi_{2,-A}+\partial_j\Phi_{3,-A}+4\partial_j\Phi_{4,-A}\\\nonumber
&-\frac{1}{2}\partial_j\Phi_{5,-A}-\frac{1}{2}\partial_j\Phi_{6,-A}+5f_0'c^4\partial_jU_{\text{\tiny EMSG},-A}\bigg\rbrace\\
&+O(c^{-4}),
\end{align}
\end{subequations}
correspond, respectively, to the Newtonian acceleration, the self-acceleration, and the $N$-body acceleration.
Here, invoking the internal dynamical equilibrium of the body, we also assume that $\frac{d}{dt}\mathfrak{P}^j_A=0$, $\frac{d}{dt}P^j_A=0$, and $\frac{d}{dt}\mathfrak{M}_A=0$.

As seen, $\big(a^j_A\big)_{\text{Newt}}$ and $\big(a^j_A\big)_{\text{N-body}}$ are entirely determined by the external components of the gravitational potentials, unaffected by internal-structure-dependent terms such as the kinetic energy $\mathcal{T}_A$ and the internal energy $E^{\text{int}}_A$. 
We have shown in \cite{2024PhRvD.110f4023N} that in the quadratic-EMSG theory, these parts of the body's acceleration ultimately reduces to the corresponding expression in GR.   
On the other hand, the asymmetry of the body apparently dictates a self-acceleration in the dynamics of the body, which is purely composed of terms dependent on the internal structure. As expected from the Strong Equivalence Principle in GR, no contribution from GR appears in $\big(\boldsymbol{a}_A\big)_{\text{self}}$; only EMSG-induced terms contribute.     
The presence of these EMSG structure-dependent terms in the equations of motion may lead to significant departures from GR, warranting careful investigation.  

To further simplify the self-interaction part of Eq.~\eqref{a_A_f}, we examine the new condition $\frac{d}{dt}\mathfrak{P}^j_A=0$ that emerges in the EMSG framework. Referring to the definition of $\mathfrak{P}^j_A$ in Eq.~\eqref{mathfrakP}, and using Eq.~\eqref{rhos}, the Newtonian limit of Eq.~\eqref{Euler-PN-EMSG}, as well as the integral identities given in Appendix C of \cite{2024PhRvD.110f4023N}, we obtain 
\begin{align}\label{vir_emsg}
\frac{d}{dt}\mathfrak{P}^j_A={\Omega_A^{**}}^j+\int_A p\,\partial_j\rho^*d^3x-\int_A{\rho^*}^2\bar{v}^j\partial_k\bar{v}^kd^3x=0.
\end{align}
This relation constitutes a brand-new equilibrium condition specific to the present theory. Imposing this condition reveals, quite remarkably, that $\big(a^j_A\big)_{\text{self}}=0$. Consequently, this modified theory, like GR, does not predict any self-acceleration. Therefore, according to this finding and the result of \cite{2024PhRvD.110f4023N}, the motion of a body within the field of an $N$-body system is governed by equations that are structurally similar in both GR and the quadratic-EMSG theory.

As a final remark in this section, it is worth noting that self-acceleration is closely related to a violation of momentum conservation \cite{will2018theory}. Keeping this fact in mind, and in light of our derivation, it is natural to anticipate the presence of a conserved PN linear momentum in the quadratic-EMSG theory, much as in GR, although it emerges through distinct internal-structure considerations. In the next section, we proceed to identify this conserved quantity explicitly.

\section{Conservation of momentum in EMSG}\label{Sec. IV}

The result obtained in the previous section naturally raises the question of the form of the conserved total momentum in the EMSG theory. 
To address this, we introduce the PN integral conservation law for the total momentum. 
There are two approaches to derive this conserved quantity. The first is to identify an energy-momentum pseudotensor, $\tau^{\mu\nu}$, whose ordinary divergence is zero in the PN limit of the quadratic-EMSG theory, i.e., $\partial_{\nu}\tau^{\mu\nu}=0$.  
This will provide us with an integral conservation law for the total momentum (to the PN order) \cite{will2018theory}. The second approach is to integrate the PN EMSG hydrodynamic equations of motion over all space and search for a time-independent quantity \cite{poisson2014gravity}. In what follows, we adopt the latter method.   

We return to the local conservation statement \eqref{nabla_T_eff} and focus on its spatial component. 
Employing the Christoffel symbols introduced in \cite{2024PhRvD.110f4023N} together with the definition of $T^{\mu\nu}_{\text{eff}}$ for a perfect fluid, and after carrying out the necessary  manipulations and simplifications, this component can be expanded up to the first PN order as follows:  
\begin{align}
\nonumber
0&=\partial_t\big(\rho^*\mu\,v^j\big)+\partial_k\big(\rho^*\mu\,v^jv^k\big)+\partial_jp-\rho^*\partial_jU+f_0'c^4\partial_j{\rho^*}^2\\\nonumber
&-\frac{1}{c^2}\bigg\lbrace\rho^*\partial_j U\Big(\Pi+\frac{3}{2}v^2-U+\frac{p}{\rho^*}\Big)-2U\partial_jp+\rho^*\partial_j\Psi\\\nonumber
&-2\rho^*\frac{d}{dt}\Big(Uv_j-2U_j\Big)-4\rho^*v^k\partial_jU_k-2f_0'c^4\Big[\partial_t\big({\rho^*}^2v^j\big)\\\nonumber
&+\partial_k\big({\rho^*}^2v^jv^k\big)-\frac{1}{2}\partial_j\Big({\rho^*}^2\big(v^2-2\,\Pi+6\,U\big)\Big)\\
&-2\rho^*\partial_jU_{\text{\tiny EMSG}}+\partial_j\big({\rho^*}^2U\big)-2{\rho^*}^2\partial_jU\Big]\bigg\rbrace+O(c^{-4}),
\end{align}
where $\mu=1+\frac{1}{c^2}\big(\Pi+\frac{p}{\rho^*}+\frac{1}{2}v^2+U\big)$. Here, the definition of the total time derivative, i.e., $d/dt=\partial_t+v^k\partial_k$, and the Newtonian limit of the EMSG Euler equation are utilized. 
Integrating the above relation over the fluid volume, we obtain
\begin{align}
\nonumber
0&=\frac{d}{dt}\bigg[\int\rho^*\mu\, v^jd^3x-\frac{1}{c^2}\bigg\lbrace 2\int \rho^*v^j\,U\,d^3x+\frac{1}{2}\int \rho^*\Phi_jd^3x\\
&-\frac{1}{2}\int \rho^*U_jd^3x-2f_0'c^4\int{\rho^*}^2v^jd^3x\bigg\rbrace\bigg],
\end{align}
after applying Gauss's theorem, the exchange trick $x\leftrightarrow x'$, and the integral identities given in Appendix C of \cite{2024PhRvD.110f4023N}. Here, to simplify this relation, we also make use of the identities $\int\rho^*\partial_jU_{\text{\tiny EMSG}}\,d^3x=-\int{\rho^*}^2\partial_j U\,d^3x$ and $\partial_{tj}X=-U_j+\Phi_j$, where $\Phi_j$ is defined in Eq.~\eqref{Phi_j}. 
Substituting the definition of $\mu$, and noting that $\int \rho^*U_j\,d^3x=\int \rho^* U\,v_j\,d^3x$, we derive the conserved linear momentum, including the EMSG contribution, as
\begin{align}
\nonumber
&P^j=\int \rho^* v^j\Big[1+\frac{1}{c^2}\Big(\Pi+\frac{1}{2}v^2-\frac{1}{2}U+\frac{p}{\rho^*}+2f_0'c^4\rho^*\Big)\Big]d^3x\\
&-\frac{1}{2c^2}\int \rho^*\Phi_jd^3x+O(c^{-4}),
\end{align}
whose total time derivative vanishes, $\frac{d}{dt}P^j=0$. 
This demonstrates that the EMSG theory admits PN integral conservation laws for the total momentum of isolated gravitating systems. Consequently, no anomalous self-acceleration arises in the EMSG framework (up to the first PN order), in agreement with the result obtained in the previous section.

\section{Discussion and Conclusion}\label{Sec. V}

A primary motivation of this work is to investigate whether self-acceleration is a feature of the EMSG theory. A challenge inherent in the matter-type modified theories of gravity, including EMSG, is the non-vanishing divergence of $T^{\mu\nu}$, which naturally opens the possibility of self-acceleration. Such self-interaction effects can, in principle, be constrained by observations of asymmetric systems, such as eccentric binary pulsars. Given their high precision, binary pulsar observations provide a stringent test of the theory's viability in the strong-gravity regime. From this perspective, investigating self-acceleration within EMSG is both relevant and necessary. 
 
The present study has followed the framework of our previous work \cite{2024PhRvD.110f4023N}, except that here we have relaxed the reflection-symmetric assumption in order to examine the role of self-acceleration in $N$-body dynamics. This relaxation comes at the cost of more intricate equations of motion within the EMSG framework. We have analyzed this relativistic effect on the motion of a massive self-gravitating body in an $N$-body system, up to the first PN order. By introducing a suitable expression for the center-of-mass acceleration, invoking several virial identities (including one that is new in the EMSG framework), and applying the resulting dynamical equilibrium conditions, we have ultimately concluded that $\boldsymbol{a}_{\text{self}}=0$.

It is worth emphasizing that this conclusion is established within the complete first PN framework, namely in the leading relativistic order where anomalous center-of-mass effects or self-acceleration could first arise in slowly moving gravitating systems. In the present analysis, the internal-structure-dependent terms appear explicitly in the intermediate steps of the derivation, see Eq.~\eqref{a_self}, but they finally cancel out after imposing the virial identities, including the new EMSG relation given in Eq.~\eqref{vir_emsg}. Therefore, the vanishing of self-acceleration is not a consequence of neglecting relevant first PN contributions, but rather a nontrivial result of the full first PN dynamics. Possible higher-order corrections beyond the first PN order remain an interesting topic for future investigation.

This noteworthy result, when combined with our previous study, demonstrates that internal-structure-dependent terms do not affect $N$-body dynamics. Consequently, the Strong Equivalence Principle\textemdash specifically its manifestation as the Gravitational Weak Equivalence Principle\textemdash is indeed satisfied in the quadratic-EMSG theory.    
Moreover, our analysis indicates that, in light of the extremely small measured self-acceleration of the center of mass in binary pulsar systems \cite{2019MNRAS.482.3249Z,2011ApJ...743..102G,2005ApJ...632.1060S,2020ApJ...898...69M}, the quadratic-EMSG theory remains viable, akin to GR, since its foremost criterion at the level considered here\textemdash consistency with current self-acceleration bounds\textemdash is satisfied.

We stress, however, that the present notion of viability refers specifically to the absence of self-acceleration in compact-body dynamics and compatibility with binary-pulsar constraints at the first PN order. A broader phenomenological assessment of the quadratic-EMSG theory should also include Solar-System tests, strong-field astrophysical observations, and cosmological data. Such constraints restrict the acceptable range of the coupling parameter $f_0'$, but do not by themselves mean that the theory is ruled out entirely. Since the EMSG corrections scale with matter variables through $T_{\mu\nu}T^{\mu\nu}$, their observational impact can depend significantly on the density regime under consideration. Determining whether a common parameter region simultaneously satisfy local tests and yields viable late-time cosmology requires a dedicated global analysis and lies beyond the scope of the present work.

An immediate consequence of the present result is that observational bounds derived from binary pulsar observations, specifically the constraints on self-acceleration (which are technically related to the second and third time derivatives of the pulsar's spin frequency in binary systems \cite{2020ApJ...898...69M,1992ApJ...393L..59W,1976ApJ...205..861W}) do not impose direct constraints on the quadratic-EMSG coupling parameter at the first PN order considered here, since the predicted self-acceleration vanishes, as in GR. This does not imply, however, that the theory is observationally unconstrained. Rather, meaningful bounds may arise from other sectors that are sensitive to matter-dependent corrections, including binary-pulsar timing parameters, compact-star structure, neutron-star mass-radius measurements, gravitational-wave inspiral phasing, and cosmological observations. Accordingly, the absence of self-acceleration in EMSG should be viewed not as a loss of predictability, but as a successful and consistent result, while the most stringent parameter constraints are expected to arise from compact-object and cosmological observations.

It is well established that self-acceleration is closely tied to the violation of momentum conservation \cite{will2018theory}. Building on this point, we have stretched our calculation and derived the PN integral conservation law for the total momentum, fulfilling the requirement of zero self-acceleration within the EMSG framework.  Quoting the result of this law, we have inferred that there is an EMSG linear momentum, $\boldsymbol{P}$, that is conserved up to the first PN order.
In the language of the PPN framework, and according to the results of Secs. \ref{Sec. III} and \ref{Sec. IV}, one can conclude that the PPN conservation-law parameters $\alpha_3$, $\zeta_1$, $\zeta_2$, $\zeta_3$, and $\zeta_4$ all vanish in the quadratic-EMSG theory\textemdash ironically, just as in GR. 
From this point, it can be asserted that this theory falls into the category of semiconservative theories \cite{will2018theory}.  
Furthermore, based on the result of \cite{2024PhRvD.110f4023N}, it follows that the quadratic-EMSG theory yields the same values for the PPN parameters $\gamma$ and $\beta$ as GR, namely $\gamma=1$ and $\beta=1$.
We leave the discussion of the remaining three PPN parameters in the EMSG theory, and more generally in matter-type modified theories, to future work. Nevertheless, the issue is not necessarily limited to these ten parameters. Since a new type of potential is added to the well-known set of PN potentials, it is plausible that additional potentials (and possibly additional PPN parameters) may be required to properly capture the PN limit of matter-type modified theories. This suggests that the PPN formalism needs to be developed for such theories. Importantly, this necessity is not unique to this theory; other modified gravity theories, such as the massive scalar-tensor theory \cite{1991ApJ...382..223H} and Chern-Simons theory \cite{2009PhR...480....1A}, also demand an extended PPN framework.

We conclude this section by discussing the conservation of energy-momentum in the EMSG theory and the status of its Lagrangian formulation. From the result of Sec. \ref{Sec. IV} and Subsec. III.B of \cite{2024PhRvD.110f4023N}, it follows that a conserved four-momentum exists in this modified theory at the PN order. Moreover, as shown in \cite{2024PhRvD.109j4055A}, the conventional formulation of this theory is confined to the level of the field equations, effectively ruling out its original Lagrangian formulation. On the other hand, according to the conjecture proposed in \cite{1974PhRvD..10.1685L} at the PN order, there is an equivalence between the existence of a conserved four-momentum in a metric theory of gravity and the existence of a Lagrangian formulation.
Thus, based on this conjecture, it is conceivable that the conventional form of the EMSG theory can be Lagrangian-based, possessing an alternative formulation that, however, cannot coincide with the original one.

\section*{acknowledgments}

We thank the anonymous referees for their useful and constructive comments.
We gratefully acknowledge the support provided by Tafresh University.

\appendix

\section{Post-Newtonian limit of EMSG}\label{app1}

In the PN approximation, we consider systems characterized by weak gravitational fields and objects moving slowly compared to the speed of light. 
The PN limit of the quadratic-EMSG theory was derived in Subsection II.B and Appendix A of \cite{2024PhRvD.110f4023N}. In this appendix, we summarize the PN relations required for the present work, in order to streamline the subsequent calculations. 
The analysis is carried out under the following assumptions:
\begin{itemize}
\item The system consists of $N$ well-separated bodies, each described as a perfect fluid governed by the PN EMSG hydrodynamic equations. Magnetic and radiation fields are neglected. 
\item Each body is self-gravitating, non-spinning, and in a state of internal dynamical equilibrium.
\item The gravitational potentials are decomposed into internal and external parts, associated with the body itself and with the remaining bodies, respectively. The internal parts are denoted by the index ``$A$", while the external ones are indicated by``$-A$". 
\item Bodies are not assumed to be symmetric about their centers of mass.  
\end{itemize}

As mentioned earlier, the matter sector of the system is modeled as a perfect fluid, with the energy-momentum tensor as 
\begin{align}
T^{\mu\nu}=\Big(\rho+\frac{\epsilon}{c^2}+\frac{p}{c^2}\Big)u^{\mu}u^{\nu}+p\,g^{\mu\nu},
\end{align}
where $\rho$ is the proper mass density, $p$ is the pressure, and $\epsilon$ is the proper internal energy density. The corresponding EMSG contribution to the energy-momentum tensor is given by  
\begin{align}
\nonumber
&T^{\mu\nu}_{\text{\tiny EMSG}}=f_0'\bigg[\Big(2c^2\rho^2+8\rho\,p+4\rho\,\epsilon+\frac{1}{c^2}\big(6p^2+8p\,\epsilon\\
&+2\epsilon^2\big)\Big)u^\mu u^\nu+\Big(c^4\rho^2+2c^2\rho\,\epsilon+3p^2+\epsilon^2\Big)g^{\mu\nu}\bigg].
\end{align}

A key result needed for the present analysis is the PN limit of the Euler equation in the quadratic-EMSG framework. The spatial component of the conservation law $\nabla_{\mu}T_{\text{eff}}^{\mu\nu}=0$ yields 
\begin{align}\label{Euler-PN-EMSG}
\nonumber
&\rho^*\frac{dv^j}{dt}=-\partial_jp+\rho^*\partial_jU+\frac{1}{c^2}\bigg\lbrace\Big(\Pi+\frac{p}{\rho^*}+\frac{1}{2}v^2+U\Big)\partial_jp\\\nonumber
&-v^j\partial_tp+\rho^*\Big[\big(v^2-4U\big)\partial_jU-v^j\big(3\partial_tU+4v^k\partial_kU\big)\\\nonumber
&+4v^k\big(\partial_kU_j-\partial_jU_k\big)+4\partial_tU_j+\partial_j\Psi\Big]\bigg\rbrace-2f_0'c^4\rho^*\bigg\lbrace\partial_j\rho^*\\\nonumber
&-\frac{1}{c^2}\Big[v^j\big(v^k\partial_k\rho^*+\rho^*\partial_kv^k\big)-\big(\Pi-\frac{3}{2}v^2-7U\big)\partial_j\rho^*\\\nonumber
&+\frac{1}{\rho^*}\partial_j(\rho^*p)+2f_0'c^4\rho^*\partial_j\rho^*-\rho^*\partial_j\big(\Pi-\frac{1}{2}v^2-3U\big)\\
&+2\partial_jU_{\text{\tiny EMSG}}\Big]\bigg\rbrace+O(c^{-4}),
\end{align}
which is expressed up to the first PN order. Here, $\Pi=\epsilon/\rho^*$. 
It should be noted that the gauge conditions, together with the continuity equation, determine the order of magnitude of the theory parameter \cite{1975PhRvD..12..376S,2023MNRAS.523.5452A}. In particular, imposing the harmonic gauge conditions has been shown to constrain the parameter of the quadratic-EMSG theory, $f_0'$, to be of the order $c^{-4}$ \cite{2024PhRvD.110f4023N}.    
Furthermore, the time component of $\nabla_{\mu}T_{\text{eff}}^{\mu\nu}=0$ leads to the local conservation of energy, i.e., the first law of thermodynamics, 
\begin{align}\label{first-law}
\frac{d\Pi}{dt}=\frac{p}{{\rho^*}^2}\frac{d\rho^*}{dt}+O(c^{-2}).
\end{align}
As expected, this relation remains unaltered in this modified theory. Importantly, local energy conservation, together with local baryon number conservation \eqref{rhos}, depends solely on the matter description (here assumed to be a perfect fluid) and not on the underlying gravitational theory \cite{will2018theory}.

In what follows, we introduce the integral form of the Newtonian and PN gravitational potentials that enter the PN Euler equation \eqref{Euler-PN-EMSG}. 
Here, $\Psi=\psi+V+\frac{1}{2}\partial_{tt} X$ and 
\begin{subequations}
\begin{align}
\label{U}
& U=G\int_{\mathcal{M}}\frac{{\rho^*}'}{\rvert{\boldsymbol{x}-\boldsymbol{x}'}\rvert}d^3x',\\
& U^j=G\int_{\mathcal{M}}\frac{{\rho^*}'v'^j}{\rvert{\boldsymbol{x}-\boldsymbol{x}'}\rvert}d^3x',\\
\label{U_EMSG}
& U_{\text{\tiny EMSG}}=G\int_{\mathcal{M}}\frac{{\rho^*}'^2}{\rvert\boldsymbol{x}-\boldsymbol{x}'\rvert}d^3x',\\
&\psi=G\int_{\mathcal{M}}\frac{{\rho^*}'}{{\rvert{\boldsymbol{x}-\boldsymbol{x}'}\rvert}}\Big(\Pi'+\frac{1}{2}v'^2-\frac{1}{2}U'\Big)d^3x',\\
& V=G\int_{\mathcal{M}}\frac{{\rho^*}'}{\rvert{\boldsymbol{x}-\boldsymbol{x}'}\rvert}\Big(\frac{3p'}{{\rho^*}'}+v'^2-\frac{1}{2}U'\Big)d^3x',\\
& X=G\int_{\mathcal{M}}{{\rho^*}'}\rvert{\boldsymbol{x}-\boldsymbol{x}'}\rvert d^3x',
\end{align}
\end{subequations}
where primed quantities are functions of $t$ and $\boldsymbol{x}'$. The integration domain $\mathcal{M}$ denotes a three-dimensional sphere of radius $\mathcal{R}$ with boundary $\partial \mathcal{M}$, representing the near-zone region in the PN framework.
As seen, a new potential\textemdash indicated by the EMSG potential $U_{\text{\tiny EMSG}}$\textemdash appears in addition to the standard set of PN potentials. This suggests that further potentials (and possibly additional PPN parameters) are required to properly capture the PN limit of matter-type modified theories. Consequently, the PPN formalism itself must be extended in order to accommodate such theories. This need for an extended PPN framework is not unique to the matter-type modified theories; other modified gravity theories, such as the massive scalar-tensor theory \cite{1991ApJ...382..223H} and Chern-Simons theory \cite{2009PhR...480....1A}, also require an extension of the PPN formalism.

\section{Force integrals}\label{app2}

In this appendix, we present the force integrals primarily to establish the framework for the calculations in Sec. \ref{Sec. III}.  To manage the complexity of Eq.~\eqref{a_j-EMSG}, we decompose it into several components, referred to as force integrals. We first introduce these forces and subsequently simplify them using the methods and mathematical techniques outlined in \cite{poisson2014gravity,2024PhRvD.110f4023N}.  

The first set of force integrals, constructed entirely within GR, is defined as 
\begin{subequations}
\begin{align}
&F_0^j\equiv\int_A\big(\rho^*\partial_jU-\partial_jp\big)\,d^3x,\\
&F_1^j\equiv\frac{1}{2c^2}\int_Av^2\partial_jp\,d^3x,\\
&F_2^j\equiv\frac{1}{c^2}\int_AU\,\partial_jp\,d^3x,\\
&F_3^j\equiv\frac{1}{c^2}\int_A\Pi\,\partial_jp\,d^3x,\\
&F_4^j\equiv\frac{1}{c^2}\int_A\frac{p}{\rho^*}\,\partial_jp\,d^3x,\\
&F_5^j\equiv-\frac{1}{c^2}\int_Av^j\,\partial_tp\,d^3x,\\
&F_6^j\equiv\frac{1}{c^2}\int_A\rho^*v^2\,\partial_jU\,d^3x,\\
&F_7^j\equiv-\frac{4}{c^2}\int_A\rho^*U\,\partial_jUd^3x,\\
&F_8^j\equiv-\frac{3}{c^2}\int_A\rho^*v^j\,\partial_tU\,d^3x,\\
&F_9^j\equiv-\frac{4}{c^2}\int_A\rho^*v^jv^k\partial_kU\,d^3x,\\
&F_{10}^j\equiv\frac{4}{c^2}\int_A\rho^*v^k\partial_kU^j\,d^3x,\\
&F_{11}^j\equiv-\frac{4}{c^2}\int_A\rho^*v^k\partial_jU_k\,d^3x,\\
&F_{12}^j\equiv\frac{4}{c^2}\int_A\rho^*\partial_tU^j\,d^3x,\\
&F_{13}^j\equiv\frac{2}{c^2}\int_A\rho^*\partial_j\Phi_1\,d^3x,\\
&F_{14}^j\equiv-\frac{1}{c^2}\int_A\rho^*\partial_j\Phi_2\,d^3x,\\
&F_{15}^j\equiv\frac{1}{c^2}\int_A\rho^*\partial_j\Phi_3\,d^3x,\\
&F_{16}^j\equiv\frac{4}{c^2}\int_A\rho^*\partial_j\Phi_4\,d^3x,\\
&F_{17}^j\equiv-\frac{1}{2c^2}\int_A\rho^*\partial_j\Phi_5\,d^3x,\\
&F_{18}^j\equiv-\frac{1}{2c^2}\int_A\rho^*\partial_j\Phi_6\,d^3x.
\end{align}
\end{subequations}
Here, the auxiliary potentials $\Phi_1$\textendash$\Phi_6$ are respectively given by 
\begin{subequations}
\begin{align}
\label{Phi_1}
&\Phi_1\equiv G\int_{\mathcal{M}}\frac{{\rho^*}'v'^2}{\rvert{\boldsymbol{x}-\boldsymbol{x}'}\rvert}d^3x',\\
&\Phi_2\equiv G\int_{\mathcal{M}}\frac{{\rho^*}'U'}{\rvert{\boldsymbol{x}-\boldsymbol{x}'}\rvert}d^3x',\\
&\Phi_3\equiv G\int_{\mathcal{M}}\frac{{\rho^*}'\Pi'}{\rvert{\boldsymbol{x}-\boldsymbol{x}'}\rvert}d^3x',\\
&\Phi_4\equiv G\int_{\mathcal{M}}\frac{p'}{\rvert{\boldsymbol{x}-\boldsymbol{x}'}\rvert}d^3x',\\
&\Phi_5\equiv G\int_{\mathcal{M}}{\rho^*}'\partial_{j'}U'\frac{\big(x-x'\big)^j}{\rvert{\boldsymbol{x}-\boldsymbol{x}'}\rvert}d^3x',\\
\label{Phi_6}
&\Phi_6\equiv G\int_{\mathcal{M}}{\rho^*}'v'_jv'_k\frac{\big(x-x'\big)^j\big(x-x'\big)^k}{\rvert{\boldsymbol{x}-\boldsymbol{x}'}\rvert^3}d^3x',\\
\label{Phi_j}
&\Phi^j\equiv G\int_{\mathcal{M}}{\rho^*}'v'_k\frac{\big(x-x'\big)^j\big(x-x'\big)^k}{\rvert{\boldsymbol{x}-\boldsymbol{x}'}\rvert^3}d^3x'.
\end{align}
\end{subequations}
To complete this set of potentials, we also introduce $\Phi_j$ here. 
The next set of force integrals, resulting from EMSG contributions, is as follows:
\begin{subequations}
\begin{align}
&F_{19}^j\equiv 3f_0'c^2\int_A\rho^*v^2\,\partial_j\rho^*\,d^3x,\\
&F_{20}^j\equiv 14f_0'c^2\int_A \rho^*U\,\partial_j\rho^*\,d^3x,\\
&F_{21}^j\equiv -2f_0'c^2\int_A \rho^*\Pi\, \partial_j\rho^*\,d^3x,\\
&F_{22}^j\equiv 2f_0'c^2\int_A \rho^*v^jv^k\partial_k\rho^*\,d^3x,\\
&F_{23}^j\equiv 2f_0'c^2\int_A {\rho^*}^2v^j\partial_kv^k\,d^3x,\\
&F_{24}^j\equiv 2f_0'c^2\int_A \partial_j\big(\rho^*p\big)\,d^3x,\\
&F_{25}^j\equiv 4{f_0'}^2c^6\int_A {\rho^*}^2\partial_j\rho^*\,d^3x,\\
&F_{26}^j\equiv f_0'c^2\int_A {\rho^*}^2\partial_jv^2\,d^3x,\\
&F_{27}^j\equiv -2f_0'c^2\int_A {\rho^*}^2\partial_j\Pi\,d^3x,\\
&F_{28}^j\equiv 6f_0'c^2\int_A {\rho^*}^2\partial_jU\,d^3x,\\
&F_{29}^j\equiv 5f_0'c^2\int_A \rho^*\partial_j U_{\text{\tiny EMSG}}\,d^3x,\\
&F_{30}^j\equiv -2f_0'c^4\int_A \rho^*\partial_j\rho^*\,d^3x.
\end{align}
\end{subequations}
Among these definitions, $F_{30}^j$ is the Newtonian term, while the others are relativistic, i.e., PN corrections. The third group of force integrals, associated with the fourth assumption discussed in Appendix \ref{app1}, is defined as  
\begin{subequations}
\begin{align}
&F_{31}^j\equiv \frac{1}{2c^2}\int_A \rho^*\bar{v}^2\partial_jU\,d^3x,\\
&F_{32}^j\equiv -\frac{1}{2c^2}\int_A\rho^*U_A\partial_jU\,d^3x,\\
&F_{33}^j\equiv \frac{1}{c^2}\int_A\rho^*\Pi\,\partial_jU\,d^3x,\\
&F_{34}^j\equiv -\frac{1}{2c^2}\int_A \bar{v}^2\partial_jp\,d^3x,\\
&F_{35}^j \equiv \frac{1}{2c^2}\int_A U_A\partial_jp\,d^3x,\\
&F_{36}^j \equiv -f_0'c^2\int_A\rho^*\bar{v}^2\partial_j\rho^*\,d^3x,\\
&F_{37}^j \equiv f_0'c^2\int_A \rho^*U_A\partial_j\rho^*\,d^3x,\\
&F_{38}^j \equiv -f_0'c^2\int_A \rho^*\partial_jp\, d^3x,
\end{align}
\end{subequations}
in which the first five integrals are GR-induced, while the last three are EMSG-induced.
 
Applying the methods described in detail in \cite{poisson2014gravity,2024PhRvD.110f4023N}, 
we derive the following expressions for the force integrals:
\begin{subequations}
\begin{align}
\label{F_0}
&F_0^j=m_A\partial_jU_{-A},\\
&F_1^j=\frac{1}{c^2}\Big[v_A^kL_A^{kj}+\frac{1}{2}\int_A\bar{v}^2\partial_jp\,d^3x\Big],\\
&F_2^j=-\frac{1}{c^2}\Big[P_A\partial_jU_{-A}+\mathcal{P}^j_A\Big],\\
&F_3^j=\frac{1}{c^2}\int_A\Pi\,\partial_jp\,d^3x,\\
&F_4^j=\frac{1}{c^2}\int_A\frac{p}{\rho^*}\,\partial_jp\,d^3x,\\
&F_5^j=-\frac{1}{c^2}\Big[v_A^j\,\frac{d}{dt}P_A+\int_A \bar{v}^j\partial_tp\,d^3x\Big],\\\nonumber
&F_6^j=\frac{1}{c^2}\Big[2\,v_A^kH_A^{kj}+t^j_A+m_Av_A^2\partial_jU_{-A}\\
&~~~~~~+2\,\mathcal{T}_A\partial_jU_{-A}\Big],\\\nonumber
&F_7^j=-\frac{4}{c^2}\Big[\Omega_A^j-2\,\Omega_A\partial_jU_{-A}+\Omega_{A}^{jk}\partial_kU_{-A}\\
&~~~~~~+m_AU_{-A}\partial_jU_{-A}\Big],\\
&F_8^j=-\frac{3}{c^2}\Big[{\mathcal{T}^*_A}^j-v_A^kH_A^{jk}+v_A^jH_A+m_Av_A^j\partial_tU_{-A}\Big],\\\nonumber
&F_9^j=-\frac{4}{c^2}\Big[v_A^jH_A+v_A^kH_A^{jk}+\mathcal{T}_A^j+m_Av_A^jv_A^k\partial_kU_{-A}\\
&~~~~~~+2\,\mathcal{T}^{jk}_A\partial_kU_{-A}\Big],\\
&F_{10}^j=\frac{4}{c^2}\Big[{\mathcal{T}^*_A}^j+v_A^jH_A-v^k_AH_A^{jk}+m_Av_A^k\partial_kU_{j,-A}\Big],\\
&F_{11}^j=-\frac{4}{c^2}m_Av_A^k\partial_jU_{k,-A},\\\nonumber
&F_{12}^j=\frac{4}{c^2}\Big[\Omega_A^j-2\,\Omega_A\partial_jU_{-A}+\mathcal{P}_A^j+v_A^kH_A^{jk}+v_A^jH_A\\
&~~~~~~+\mathcal{T}^j_A+m_A\partial_tU_{j,-A}\Big]-8f_0'c^2\Lambda^j_A,\\
&F_{13}^j=-\frac{2}{c^2}\Big[2\,v^k_AH^{kj}_A+t^j_A-m_A\partial_j\Phi_{1,-A}\Big],\\
&F_{14}^j=\frac{1}{c^2}\Big[\Omega^j_A-m_A\partial_j\Phi_{2,-A}+\Omega^{jk}_A\partial_kU_{-A}\Big],\\
&F_{15}^j=-\frac{1}{c^2}\Big[\varepsilon_A^j-m_A\partial_j\Phi_{3,-A}\Big],\\
&F_{16}^j=-\frac{4}{c^2}\Big[\mathcal{P}^j_A-m_A\partial_j\Phi_{4,-A}\Big],\\\nonumber
&F_{17}^j=-\frac{1}{2c^2}\Big[2\,\Omega^{jk}_A\partial_kU_{-A}-2\,\Omega_A\partial_jU_{-A}+\Omega^j_A+{\Omega^*_A}^j\\
&~~~~~~+m_A\partial_j\Phi_{5,-A}\Big],\\\nonumber
&F_{18}^j=-\frac{1}{c^2}\Big[v^j_AH_A+v_A^kH_A^{jk}+\mathcal{T}^j_A-3\,v_A^kK_A^{jk}-\frac{3}{2}{\mathcal{T}^{**}_A}^j\\
&~~~~~~+\frac{1}{2}m_A\partial_j\Phi_{6,-A}\Big],
\end{align}
\end{subequations}
as well as
\begin{subequations}
\begin{align}
&F_{19}^j=3f_0'c^2\Big[2\,v_A^kQ_A^{kj}+{Q^*_A}^j\Big],\\
&F_{20}^j=14f_0'c^2\Big[\Lambda_A^j-\frac{1}{2}\mathfrak{M}_A\partial_jU_{-A}\Big],\\
&F_{21}^j=-2f_0'c^2\int_A\rho^*\Pi\,\partial_j\rho^*\,d^3x,\\
&F_{22}^j=2f_0'c^2\Big[v_A^kQ_A^{jk}+Q_A^j+v_A^jQ_A\Big],\\
&F_{23}^j=2f_0'c^2\Big[{\Omega^{**}_A}^j-\frac{d}{dt}\mathfrak{P}^j_A+\int_Ap\,\partial_j\rho^*\,d^3x-2\,v_A^jQ_A\Big],\\
&F_{24}^j=0,\\
&F_{25}^j=0,\\
&F_{26}^j=-2f_0'c^2\Big[{Q^*_A}^j+2\,v_A^kQ_A^{kj}\Big],\\
&F_{27}^j=4f_0'c^2\int_A\rho^*\Pi\,\partial_j\rho^*\,d^3x,\\
&F_{28}^j=6f_0'c^2\Big[{\Omega^{**}_A}^j+\mathfrak{M}_A\partial_jU_{-A}\Big],\\
&F_{29}^j=-5f_0'c^2\Big[{\Omega^{**}_A}^j-m_A\partial_jU_{\text{\tiny EMSG},-A}\Big],\\
&F_{30}^j=0,\\
&F_{31}^j=\frac{1}{2c^2}\Big[t_A^j+2\,\mathcal{T}_A\partial_jU_{-A}\Big],\\
&F_{32}^j=-\frac{1}{2c^2}\Big[\Omega^j_A-2\,\Omega_A\partial_jU_{-A}\Big],\\
&F_{33}^j=\frac{1}{c^2}\Big[\varepsilon^j_A+E_A^{\text{int}}\partial_jU_{-A}\Big],\\
&F_{34}^j=-\frac{1}{2c^2}\int_A\bar{v}^2\,\partial_jp\,d^3x,\\
&F_{35}^j=-\frac{1}{2c^2}\mathcal{P}^j_A,\\
&F_{36}^j=-f_0'c^2{Q^*_A}^j,\\
&F_{37}^j=f_0'c^2\Lambda^j_A,\\
\label{F_38}
&F_{38}^j = -f_0'c^2\int_A \rho^*\partial_jp\, d^3x.
\end{align}
\end{subequations}
In these integrals, the external parts of all potentials, after differentiation, are evaluated at the leading order of the center-of-mass position, i.e., $\boldsymbol{x}=\boldsymbol{r}_{A(0)}$. For instance,  $\partial_jU_{-A}=\partial_jU_{-A}(t,\boldsymbol{r}_{A(0)})$. Note that certain integral terms in the above forces are left unsimplified, since they ultimately cancel each other out in the final expression of the acceleration.
   
We define below the scalar, vector, and tensor quantities that appear in the force integrals. Each type of quantity is introduced separately for clarity. The scalar quantities are defined as follows:   
\begin{subequations}
\begin{align}
\label{eq1}
&\mathcal{T}_A\equiv \frac{1}{2}\int_A \rho^*v^2\,d^3x,\\
&\Omega_A\equiv -\frac{1}{2}G\int_A\frac{\rho^*{\rho^*}'}{\rvert{\boldsymbol{x}-\boldsymbol{x}'}\rvert}d^3x'd^3x,\\
&E_A^{\text{int}}\equiv \int_A \rho^*\Pi\,d^3x,\\
\label{eq4}
&\mathfrak{M}_A\equiv \int_A {\rho^*}^2\,d^3x\\
&Q_A\equiv\int_A\rho^*\bar{v}^k\partial_k\rho^*\,d^3x,\\
&P_A\equiv \int_Ap\,d^3x,\\
&H_A\equiv G\int_A\rho^*{\rho^*}'\frac{\bar{\boldsymbol{v}}'\cdot\big(\boldsymbol{x}-\boldsymbol{x}'\big)}{\lvert\boldsymbol{x}-\boldsymbol{x}'\rvert^3}\,d^3x'd^3x.
\end{align}
\end{subequations} 
The vector quantities are defined as:  
\begin{subequations}
\begin{align}
&t_A^j\equiv G\int_A\rho^*{\rho^*}'\frac{\bar{v}'^2\big(x-x'\big)^j}{\rvert{\boldsymbol{x}-\boldsymbol{x}'}\rvert^3}d^3x'd^3x,\\
\label{T_A}
&\mathcal{T}^j_A\equiv G\int_A\rho^*{\rho^*}'\frac{\bar{v}'^j\bar{\boldsymbol{v}}'\cdot\big(\boldsymbol{x}-\boldsymbol{x}'\big)}{\rvert{\boldsymbol{x}-\boldsymbol{x}'}\rvert^3}d^3x'd^3x,\\
\label{T*_A}
&{\mathcal{T}^*_A}^j\equiv G\int_A\rho^*{\rho^*}'\frac{\bar{v}^j\bar{\boldsymbol{v}}'\cdot\big(\boldsymbol{x}-\boldsymbol{x}'\big)}{\rvert{\boldsymbol{x}-\boldsymbol{x}'}\rvert^3}d^3x'd^3x,\\
&{\mathcal{T}^{**}_A}^j\equiv G\int_A\rho^*{\rho^*}'\frac{(x-x'\big)^j\big[\bar{\boldsymbol{v}}'\cdot\big(\boldsymbol{x}-\boldsymbol{x}'\big)\big]^2}{\rvert{\boldsymbol{x}-\boldsymbol{x}'}\rvert^5}d^3x'd^3x,\\
&\Omega^j_A\equiv G^2\int_A\rho^*{\rho^*}'{\rho^*}''\frac{\big(x-x'\big)^j}{\rvert{\boldsymbol{x}'-\boldsymbol{x}''}\rvert\rvert\boldsymbol{x}-\boldsymbol{x}'\rvert^3}d^3x''d^3x'd^3x,\\\nonumber
&{\Omega^*_A}^j\equiv G^2\int_A\rho^*{\rho^*}'{\rho^*}''\big(\boldsymbol{x}'-\boldsymbol{x}''\big)\cdot\big(\boldsymbol{x}-\boldsymbol{x}'\big)\\
\label{Omega*_A}
&\times\frac{\big(x-x'\big)^j}{\lvert{\boldsymbol{x}'-\boldsymbol{x}''}\rvert^3\lvert\boldsymbol{x}-\boldsymbol{x}'\rvert^3}d^3x''d^3x'd^3x,\\
\label{Omega**_A}
&{\Omega^{**}_A}^j\equiv G\int_A\rho^*{\rho^*}'^2\frac{\big(x-x'\big)^j}{\rvert{\boldsymbol{x}-\boldsymbol{x}'}\rvert^3}d^3x'd^3x,\\
\label{P_A}
&\mathcal{P}^j_A\equiv  G\int_A\rho^*p'\frac{\big(x-x'\big)^j}{\lvert\boldsymbol{x}-\boldsymbol{x}'\rvert^3}d^3x'd^3x,\\
&\varepsilon^j_A\equiv G\int_A\rho^*{\rho^*}'\Pi'\frac{\big(x-x'\big)^j}{\lvert\boldsymbol{x}-\boldsymbol{x}'\rvert^3}d^3x'd^3x,\\
&\Lambda^j_A\equiv G\int_A \rho^*{\rho^*}'\frac{\partial_{j'}{\rho^*}'}{\lvert\boldsymbol{x}-\boldsymbol{x}'\rvert}d^3x'd^3x,\\
\label{mathfrakP}
&\mathfrak{P}^j_A\equiv \int_A{\rho^*}^2\bar{v}^j\,d^3x,\\
&Q_A^j\equiv\int_A\rho^*\bar{v}^j\bar{v}^k\partial_k\rho^*\,d^3x,\\
&{Q_A^{*}}^j\equiv\int_A\rho^*\bar{v}^2\partial_j\rho^*\,d^3x,
\end{align}
\end{subequations}
and the tensor quantities are defined as:
\begin{subequations}
\begin{align}
\label{I^jk}
&I^{jk}_A\equiv \int_A\rho^*\bar{x}^j\bar{x}^k\,d^3x,\\
&S^{jk}_A\equiv \int_A\rho^*\big(\bar{x}^j\bar{v}^k-\bar{x}^k\bar{v}^j\big)d^3x,\\
&\mathcal{T}^{jk}_A\equiv \frac{1}{2}\int_A\rho^*\bar{v}^j\bar{v}^k\,d^3x,\\
& L_A^{jk}\equiv \int_A\bar{v}^j\partial_k p\,d^3x,\\
&\Omega^{jk}_A\equiv -\frac{1}{2}G\int_A\rho^*{\rho^*}'\frac{\big(x-x'\big)^j\big(x-x'\big)^k}{\lvert\boldsymbol{x}-\boldsymbol{x}'\rvert^3}d^3x'd^3x,\\
&H^{jk}_A\equiv G\int_A\rho^*{\rho^*}'\frac{\bar{v}'^j\big(x-x'\big)^k}{\lvert\boldsymbol{x}-\boldsymbol{x}'\rvert^3}d^3x'd^3x,\\
&K^{jk}_A\equiv G\int_A\rho^*{\rho^*}'\frac{\bar{\boldsymbol{v}}'\cdot\big(\boldsymbol{x}-\boldsymbol{x}'\big)\big(x-x'\big)^j\big(x-x'\big)^k}{\lvert\boldsymbol{x}-\boldsymbol{x}'\rvert^5}d^3x'd^3x,\\
&Q^{jk}_A\equiv \int_A\rho^*\bar{v}^j\partial_k\rho^*\,d^3x.
\end{align}
\end{subequations}
Among these, $\mathfrak{M}$, $Q$, $Q^j$, ${Q^{*}}^j$, $\mathfrak{P}^j$, $\Lambda^j$, ${\Omega^{**}}^j$, and $Q^{jk}$ are new quantities that arise within the quadratic-EMSG framework.
It can be demonstrated that $Q_A=Q_A^{kk}=\frac{1}{2}\frac{d\mathfrak{M}_A}{dt}$ and $\int_A{\rho^*}^2\bar{v}^k\partial_j\bar{v}_k\,d^3x=-{Q_A^{*}}^j$. These relations are used to simplify the final form of the acceleration. 
  
\bibliographystyle{apsrev4-1}
\bibliography{short,conservation_of_mass_center_motion}

\end{document}